\documentclass[aps,pra,groupedaddress]{revtex4}

\usepackage{graphicx}
\begin{document}


\title{Entanglement and four wave mixing effects in
the dissipation free nonlinear interaction of two photons
at a single atom}

\author{Holger F. Hofmann $^{1,2}$}
\email{h.hofmann@osa.org}
\author{Kunihiro Kojima $^2$}
\author{Shigeki Takeuchi $^{1,2}$}
\author{Keiji Sasaki$^2$}
\affiliation{$^1$ PRESTO, 
Japan Science and Technology Corporation 
(JST)\\
$^2$ Research Institute for Electronic Science, Hokkaido 
University\\
Kita-12 Nishi-6, Kita-ku, Sapporo 060-0812, Japan}

\date{\today}

\begin{abstract}
We investigate the nonlinear interaction between two photons
in a single input pulse at an atomic two level nonlinearity. 
A one dimensional model for the propagation of light to and 
from the atom is used to describe the precise spatiotemporal 
coherence of the two photon state. It is shown that the 
interaction generates spatiotemporal entanglement in the
output state similar to the entanglement observed in parametric
downconversion. A method of generating photon pairs from 
coherent pump light using this quantum mechanical four wave 
mixing process is proposed.
\end{abstract}


\maketitle

\section{Introduction}
Optical nonlinearities sensitive to individual photons
may provide interesting new possibilities of controlling
and manipulating the quantum states of light \cite{Poi93,Tur95,
Ima97,Tho98,Wer99,Reb99,Hof00a,Hof00b,Fos00,Res02}.
Possible applications of such nonlinearities include 
quantum nondemolition measurements of photon number \cite{Wal}
and quantum logic circuits for photonic qubits \cite{Nie}.
Experimentally, sufficiently strong nonlinearities have been
achieved in cavity quantum electrodynamics, where cavity
confinement can enhance the coupling between a single two
level atom and the input field \cite{Tur95}. By optimizing
the suppression of uncontrollable photon losses in such
systems, it may be possible to realize a fully quantum coherent
photon-photon interaction \cite{Hof02}. 
The analysis of such a quantum level nonlinearity then requires
a quantum mechanical treatment of the spatiotemporal 
coherence in the input and output fields. Specifically, 
spontaneous four wave mixing effects may entangle the two 
input photons in their spatial coordinates. This entanglement 
appears to introduce noise in the single photon coherence, 
even though the two photons are still in a quantum mechanically
pure state.
 
In order to investigate such effects, we apply a one dimensional
model of light field propagation to and from a single two level
atom \cite{Hof95, Koj02}. If photon losses are avoided, it is then 
possible to determine the response functions for single photon 
and for two photon inputs. Using these response functions, we 
derive the output state for a resonant rectangular input. 
We discuss the implications of this result for coherent 
input fields and show that it is possible to create entangled
photon pairs from coherent input light by using an interferometric
strategy similar to the one recently applied in parametric 
downconversion \cite{Lu02}. 

\section{One dimensional model of light field propagation}
If the transversal beam profile is known, it is sufficient to
describe the propagation of light to and from a system using
only a single spatial coordinate. 
In free space, the propagation velocity $c$ is constant. 
The linear propagation process can then be described by a 
dispersion relation of $\omega = c k$, where $k$ is a
scalar \cite{Hof95}.
If this approximation is applied to the interaction of
electromagnetic field with a single two level system, 
the transversal profile of the k-space eigenmodes is defined 
by the coupling characteristics of the two level system 
to the three dimensional field in free space. 
As has been discussed in \cite{Hof95}, the single spatial
coordinate $r$ corresponding to the wavevector $k$ then
represents the distance from the system at $r=0$, 
where negative values indicate propagation towards the 
system and positive values indicate propagation away from 
the system.  

\begin{figure}
\begin{picture}(300,200)
\thicklines
\put(20,50){\vector(1,0){260}}
\put(275,30){\makebox(20,20){\large $r$}}
\put(150.5,50){\circle*{8}}
\put(150,40){\line(0,1){6}}
\put(140,20){\makebox(20,20){\large $r=0$}}
\put(80,25){\makebox(20,20){\large $r<0$}}
\put(200,25){\makebox(20,20){\large $r>0$}}
\put(150,95){\vector(0,-1){40}}
\put(160,80){\makebox(60,15){Two level atom}}

\bezier{300}(150,95)(167,95)(178,107)
\bezier{300}(178,107)(190,118)(190,135)
\bezier{300}(190,135)(190,152)(178,163)
\bezier{300}(178,163)(167,175)(150,175)
\bezier{300}(150,175)(133,175)(122,163)
\bezier{300}(122,163)(110,152)(110,135)
\bezier{300}(110,135)(110,118)(122,107)
\bezier{300}(122,107)(133,95)(150,95)

\put(125,115){\line(1,0){50}}
\put(125,155){\line(1,0){50}}
\put(135,100){\makebox(30,15){$\mid \! G \rangle$}}
\put(135,155){\makebox(30,15){$\mid \! E \rangle$}}

\put(150,118){\line(0,1){34}}
\put(150,118){\line(1,2){5}}
\put(150,118){\line(-1,2){5}}
\put(150,152){\line(1,-2){5}}
\put(150,152){\line(-1,-2){5}}

\end{picture}
\caption{\label{schematic} Schematic representation of
the one dimensional model for light field propagation in
the field-atom interaction. There is only one direction of
propagation. $r<0$ represents light propagating towards 
the atom and $r>0$ represents light propagating away from the
atom. The interaction takes place locally at $r=0$.}
\end{figure}
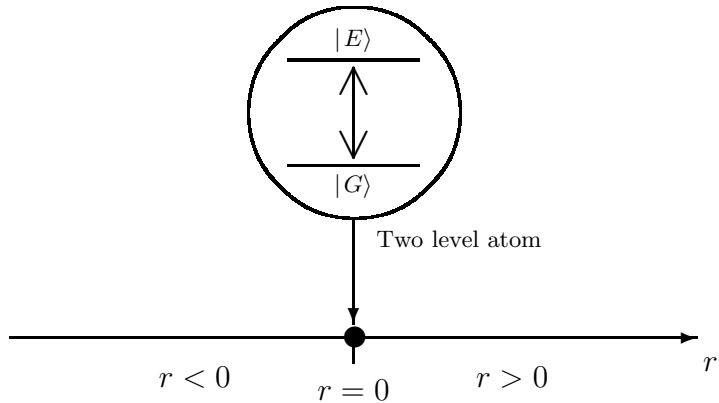

Figure \ref{schematic} shows a schematic representation of
the model. The Hamiltonian of this system can be written as
\begin{eqnarray}
\label{eq:Hamilton}
\hat{H}_{1D} &=& \hat{H}_{\mbox{prop.}} + \hat{H}_{\mbox{abs.}}  
\nonumber \\
\mbox{with} && \hat{H}_{\mbox{prop.}}
= \int dk \; \hbar c k \; \hat{b}_k^\dagger \hat{b}_k,
\nonumber \\
            && \hat{H}_{\mbox{abs.}}
= \int dk \; \hbar \sqrt{\frac{c\Gamma}{\pi}} 
(\hat{b}_k^\dagger \hat{\sigma}_-
+\hat{\sigma}_-^\dagger \hat{b}_k),
\end{eqnarray}
where $\hat{b}_k$ is the photon annihilation operator in 
$k$-space, and $\sigma_- = \mid G \rangle\langle E \mid$ is the
atomic annihilation operator describing coherence between
the ground state $\mid G \rangle$ and the excited state
$\mid E \rangle$. The coupling strength is expressed in terms
of the dipole relaxation rate $\Gamma$. This rate defines the 
characteristic timescale of the coherent interaction between
the light field and the two level atom. For convenience, the
resonant frequency of the atom has been set to zero. 
Note that this merely corresponds to a rotating frame of 
reference for the phase oscillations, so that all frequencies
are expressed as frequency shifts relative to the resonant
frequency $\omega_0 = c k_0$. 

Experimentally, the model presented here could be realized
using a one sided microcavity \cite{Hof02,Koj02}. 
Losses to transversal 
light field modes can then be minimized and almost all the
light emitted by a two level atom inside the cavity is
emitted along the axis of the cavity. If this ideal condition
cannot be met, the model described here could still be applied.
However, it would be necessary to treat the losses as a 
transversal mode mismatch between the input and output beams 
and the one dimensional field actually interacting with the
single atom.

\section{Local absorption and emission}

It is now possible to formulate the Schr\"odinger equations
for the single photon case by defining the one photon basis
as $\mid k \rangle$ for one photon in a $k$-eigenstate, and
$\mid E \rangle$ for the excited atom with no photon in free
space. The quantum state $\mid \psi (t) \rangle$ is then 
described by the components 
$\psi(k;t)=\langle k \mid \psi(t)\rangle$ and
$\psi(E;t)=\langle E \mid \psi(t)\rangle$. 
The temporal dynamics of these components is given by
\begin{eqnarray}
\frac{d}{dt} \psi(k;t) &=& 
-i ck \psi(k;t) -i \sqrt{\frac{c\Gamma}{\pi}} \psi(E;t)
\nonumber \\
\frac{d}{dt} \psi(E;t) &=&
-i \sqrt{\frac{c\Gamma}{\pi}} \int dk \; \psi(k;t). 
\end{eqnarray}
These equations of motion can now be transformed into 
real space coordinates $r$ by using the Fourier transform
\begin{equation}
\psi(r;t) = \frac{1}{\sqrt{2\pi}}
\int dk \; \exp(i kr) \psi(k;t).
\end{equation}
The equation for the propagating field then reads
\begin{equation}
\label{eq:prop}
\frac{d}{dt} \psi(r;t) = 
-c \frac{\partial}{\partial r} \psi(r;t) 
-i \sqrt{2 c\Gamma} \; \delta(r) \; \psi(E;t).
\end{equation}
As a result of the integration over $k$, this equation of
motion now includes a delta function expressing the 
locality of emission. Since the time evolution should be 
continuous, this delta function implies a jump of $\psi(r;t)$
at $r=0$. By integrating equation (\ref{eq:prop}), the
discontinuity is found to be given by
\begin{equation}
\label{eq:jump}
\psi(r\to +0;t)-\psi(r\to -0;t)
= -i \sqrt{\frac{2\Gamma}{c}} \; \psi(E;t).
\end{equation}
Emission and absorption are therefore described by the 
instantaneous addition of an amplitude proportional to 
$\psi(E;t)$ to the single photon wavefunction
propagating from $r<0$ to $r>0$.
At $r\neq 0$, the dynamics of $\psi(r;t)$ is simply 
described by linear propagation, $\psi(r;t)=\psi(r-ct;0)$.

In order to obtain the dynamics of $\psi(E;t)$, it is 
necessary to define the integral corresponding to 
$\psi(r=0;t)$. The proper result is obtained by taking
the average of the incoming amplitude $\psi(r\to -0;t)$ 
and the outgoing amplitude $\psi(r\to +0;t)$.
However, it is convenient to use the result of equation
(\ref{eq:jump}) to express the dynamics of $\psi(E;t)$
entirely in terms of the incoming amplitude $\psi(r\to -0;t)$.
It then reads
\begin{equation}
\label{eq:atom}
\frac{d}{dt} \psi(E;t) = - \Gamma \psi(E;t)
-i \sqrt{2 c\Gamma} \psi(r\to -0;t).
\end{equation}
The amplitude of the excited state $\psi(E;t)$ can therefore
be obtained from an integration of the previous incoming
field amplitudes $\psi(r\to -0;t)$. Since the dynamics
of these amplitudes are given by linear propagation at a
constant velocity of $c$, they can be obtained from the
initial single photon wavefunction at $r<0$ using the 
linear propagation dynamics mentioned above.

With these results, it is possible to integrate the 
equations of motion from any initial time $t_{\mbox{\small in}}$
to any final time $t_{\mbox{\small out}}$. In particular, the 
output field within $0<r<c (t_{\mbox{\small out}}-t_{\mbox{\small in}})$
for $\psi(E;t_{\mbox{\small in}})=0$ is given by
\begin{eqnarray}
\label{eq:integral}
\psi(r;t_{\mbox{out}}) &=& 
\psi(r-c (t_{\mbox{out}}-t_{\mbox{\small in}});t_{\mbox{\small in}})
-i \sqrt{\frac{2\Gamma}{c}} \psi(E;t_{\mbox{\small out}}-r/c)
\nonumber 
\\
&=&
\psi(r-c (t_{\mbox{\small out}}-t_{\mbox{\small in}});t_{\mbox{\small in}})
\nonumber
\\ && \hspace{0.5cm}
-2 \frac{\Gamma}{c} 
\int_{r-c (t_{\mbox{\small out}}-t_{\mbox{\small in}})}^{0} 
\hspace{-1.2cm} dr^\prime \hspace{0.7cm}
\exp \left(-\frac{\Gamma}{c}(r-r^\prime-
c (t_{\mbox{\small out}}-t_{\mbox{\small in}})\right)\;
\psi(r^\prime;t_{\mbox{\small in}}).
\end{eqnarray}
As the first line of equation (\ref{eq:integral}) shows, the
output wavefunction is a superposition of a component that 
propagated past the atom unchanged and a component emitted by 
the excited atom. Since the atom was initially in the ground
state, the emission can be traced to absorptions of the incoming
wavefunction, as represented by the integral in the last 
line of equation (\ref{eq:integral}).
The output wavefunction at $r>0$ can thus be represented as a 
linear function of the input wavefunction at $r<0$.

\section{Many photon effects}

The advantage of a local description of the field-atom
interaction is that it is easily extended to multiple 
photons. No matter how high the photon density is,
we can always define a region from $r=-\epsilon$ to
$r=+\epsilon$ around the atom small enough to contain
only one photon. In order to solve the field-atom 
interaction problem for many photons, it is therefore 
only necessary to consider what happens if a photon
interacts with the excited atom. 

For this purpose, it is useful to define the many photon
Hilbert space as a product space of independent particles.
The bosonic nature of photons must then be included in the
symmetry of the initial state. For reasons of consistency,
it is then also necessary to distinguish the origin of an
excitation, effectively treating the excited state as a
state of the photon. The two photon wavefunction is then
given by the amplitudes for two photons in free space,
$\psi(r_1,r_2)$, the amplitudes for one photon in free space
and one photon at the atom, $\psi(r_1,E)$ or $\psi(E,r_2)$,
and the amplitude for a double excitation, $\psi(E,E)$.
For a two level atom, the latter must always be zero.
In the Hamiltonian given by equation (\ref{eq:Hamilton}),
this fact is expressed by the difference between the atomic
annihilation operator $\hat{\sigma}_-$ and the annihilation
operators of harmonic oscillators. Within the product space
of independent particles, this difference is simply represented
by setting the matrix elements between single excitation and
double excitation to zero. The Schr\"odinger equation for 
the two photon wavefunction then reads
\begin{eqnarray}
\label{eq:twopse}
\frac{d}{dt} \psi(r_1,r_2;t) &=& 
-c \frac{\partial}{\partial r_1} \psi(r_1,r_2;t) 
-i \sqrt{2 c\Gamma} \; \delta(r_1) \; \psi(E,r_2;t)
\nonumber \\ &&
-c \frac{\partial}{\partial r_2} \psi(r_1,r_2;t) 
-i \sqrt{2 c\Gamma} \; \delta(r_2) \; \psi(r_1,E;t)
\nonumber \\ 
\frac{d}{dt} \psi(E,r_2;t) &=& - \Gamma \psi(E,r_2;t)
-i \sqrt{2 c\Gamma} \psi(r_1\to -0,r_2;t)
\nonumber \\ &&
-c \frac{\partial}{\partial r_2} \psi(E,r_2;t) 
\hspace{1cm}\mbox{(*)}
\nonumber \\ 
\frac{d}{dt} \psi(r_1,E;t) &=& 
-c \frac{\partial}{\partial r_1} \psi(r_1,E;t) 
\hspace{1cm}\mbox{(*)}
\nonumber \\ &&
- \Gamma \psi(r_1,E;t)
-i \sqrt{2 c\Gamma} \psi(r_1,r_2\to -0;t)
,
\end{eqnarray}
where (*) marks the missing two photon absorption terms. 
This two photon Schr\"odinger equation describes
the nearly independent dynamics of two separate photons,
except for the absence of absorption for one photon if
the other photon has been absorbed by the atom. 
The integration of the two photon Schr\"odinger equation
can therefore be achieved by using the single photon 
results and setting all contributions of double
excitation to zero \cite{Koj02}. In the following, however, 
we will present an alternative solution of the dynamics 
based on the two photon interaction represented by the 
missing double excitation terms in equation (\ref{eq:twopse}).
This procedure has the advantage that it can be easily 
extended to three or more photons and may therefore provide
a useful foundation for further investigations.

\section{Single photon and two photon response functions}
Using the results for local emission and absorption,
it is possible to evaluate the effects of the atom-field
interaction on an arbitrary single photon wavefunction. 
For this purpose, it is useful to define
a time independent characterization of the input and
output wavefunction. In the context of our model, this
characterization is easy to obtain since the propagation
before and after the interaction processes does not change
the shape of the wavepacket. For the single photon cases, 
the input and output wavefunctions can therefore be given by
\begin{eqnarray}
\psi_{\mbox{in}}(x) &=& 
\lim_{t_{\mbox{\small in}}\to -\infty} 
\psi(r=x+c \, t_{\mbox{\small in}};t_{\mbox{\small in}})
\nonumber \\
\psi_{\mbox{out}}(x) &=& 
\lim_{t_{\mbox{\small out}}\to +\infty} 
\psi(r=x+c \, t_{\mbox{\small out}};t_{\mbox{\small out}}).
\end{eqnarray}
According to equation (\ref{eq:integral}), the output wavefunction
can be obtained from the input wavefunction using a linear
response function $U_{1}(x;x^\prime)$ such that
\begin{equation}
\psi_{\mbox{out}}(x) =
\int_{-\infty}^{\infty} dx^\prime U_1(x;x^\prime) 
\psi_{\mbox{in}}(x^\prime). 
\end{equation}
The single photon response function reads
\begin{equation}
\label{eq:U1}
U_{1}(x;x^\prime) = \left \{ 
\begin{array}{cl}
\delta(x^\prime-x) - 2\frac{\Gamma}{c} 
\exp\left(-\frac{\Gamma}{c}(x^\prime-x)\right) & 
\mbox{for}\hspace{0.2cm} x \leq x^\prime
\\[0.1cm]
0 & \mbox{for}\hspace{0.2cm} x > x^\prime.
\end{array}
\right.
\end{equation}
Note that the response function $U_{1}(x;x^\prime)$ is
a representation of the unitary operation describing the
time evolution of the field-atom interaction. It therefore
preserves the norm of the wavefunction given by the 
integral over the absolute square.

Likewise, the field-atom interaction of a two photon
wavefunction can be described by a linear response formalism.
The input and output wavefunctions are then described by
\begin{eqnarray}
\psi_{\mbox{in}}(x_1,x_2) &=& 
\lim_{t_{\mbox{\small in}}\to -\infty} \psi(r_1=x_1+c \, 
t_{\mbox{\small in}},
r_2=x_2+ct_{\mbox{in}};t_{\mbox{\small in}})
\nonumber \\
\psi_{\mbox{out}}(x_1,x_2) &=& 
\lim_{t_{\mbox{\small out}}\to +\infty} \psi(r_1=x_1+ct_{\mbox{out}},
r_2=x_2+c \, t_{\mbox{\small out}};t_{\mbox{\small out}}).
\end{eqnarray}
The unitary transform of the input state into the output
state can also be described by linear response function,
\begin{equation}
\label{eq:2pU}
\psi_{\mbox{out}}(x_1,x_2) =
\int_{-\infty}^{\infty} dx_1^\prime dx_2^\prime 
U_2(x_1,x_2;x_1^\prime,x_2^\prime) 
\psi_{\mbox{in}}(x_1^\prime,x_2^\prime). 
\end{equation}
If the two photons are always very far apart 
$(x_1-x_2\gg \Gamma/c)$, or if the atom is replaced with a
harmonic oscillator, the propagation of the two photons
must be independent of each other. In this case, the
response function is equal to the product of two single
photon response functions,
\begin{equation}
U_{\mbox{lin.}}(x_1,x_2;x_1^\prime,x_2^\prime)
= U_1(x_1;x_1^\prime) U_1(x_2;x_2^\prime).  
\end{equation}
This response function corresponds to the linear part
of the field-atom interaction.
However, the absence of two photon absorption in the
dynamics causes a coupling between the photons that can
be described by a nonlinear correction $\Delta U_{\mbox{nonlin.}}$
such that
\begin{equation}
\label{eq:DU}
U_2(x_1,x_2;x_1^\prime,x_2^\prime)
=  U_1(x_1;x_1^\prime) U_1(x_2;x_2^\prime)
+ \Delta U_{\mbox{nonlin.}}(x_1,x_2;x_1^\prime,x_2^\prime).
\end{equation}
According to the considerations in the previous section,
$\Delta U_{\mbox{nonlin.}}$ can be found by integrating the
contributions from double excitations in $U_{\mbox{lin.}}$.
The result reads
\begin{equation}
\Delta U_{\mbox{nonlin.}}(x_1,x_2;x_1^\prime,x_2^\prime) 
= \left \{ 
\begin{array}{cl}
- 4\frac{\Gamma^2}{c^2} 
\exp\left(-\frac{\Gamma}{c}(x_1^\prime-x_1)\right)
\exp\left(-\frac{\Gamma}{c}(x_2^\prime-x_2)\right) & 
\mbox{for}\hspace{0.2cm} \mbox{Max}\{x_1,x_2\} < 
\mbox{Min}\{x_1^\prime,x_2^\prime\}
\\[0.2cm]
0 & \mbox{else},
\end{array}
\right. 
\end{equation}
where the minimum $\mbox{Min}\{x_1^\prime,x_2^\prime\}$ 
effectively defines the latest absorption time and
the maximum $\mbox{Max}\{x_1,x_2\}$ defines the earliest 
emission. Thus the nonlinearity removes all components
where the first emission occurs only after the second 
absorption \cite{Koj02}.  

It is now possible to derive the output wavefunction
for any two photon input wavefunction by integrating
equation (\ref{eq:2pU}) using the expressions for
$U_1$ and for $\Delta U_{\mbox{nonlin.}}$ given by
equations (\ref{eq:U1}) and (\ref{eq:DU}), respectively.
If the input is a single mode two photon pulse, the
input wavefunction can be written as a product state
\begin{equation}
\psi_{\mbox{in}}(x_1,x_2) = \phi_{\mbox{in}}(x_1) 
\phi_{\mbox{in}}(x_2), 
\end{equation}
where $\phi_{\mbox{in}}$ defines the shape of the input pulse.
The quantum state of the output field can then be 
described by 
\begin{equation}
\psi_{\mbox{out}}(x_1,x_2) = 
\phi_{\mbox{out}}(x_1) \phi_{\mbox{out}}(x_2)
+ \Delta \psi_{\mbox{nonlin.}}(x_1,x_2),
\end{equation}
where $\phi_{\mbox{out}}$ describes the linear single 
photon response given by
\begin{equation}
\phi_{\mbox{out}}(x) =
\int_{-\infty}^{\infty} dx^\prime U_1(x;x^\prime) 
\phi_{\mbox{in}}(x^\prime), 
\end{equation}
and the nonlinear contribution is directly obtained from
\begin{equation}
\label{eq:DUresp}
\Delta \psi_{\mbox{nonlin.}}(x_1,x_2) =
\int_{-\infty}^{\infty} dx_1^\prime dx_2^\prime 
\Delta U_{\mbox{nonlin.}}(x_1,x_2;x_1^\prime,x_2^\prime) 
\phi_{\mbox{in}}(x_1^\prime)\phi_{\mbox{in}}(x_2^\prime). 
\end{equation}
These equations describe the nonlinear response of the two
level atom at the quantum level. It is now possible to 
apply this response function to a variety of input states.
In the following, we will focus on the case of a
resonant rectangular wavepacket. 

\section{The quantum level nonlinearity at resonance}

\begin{figure}
\begin{picture}(500,260)
\put(10,0){\makebox(240,240){\includegraphics{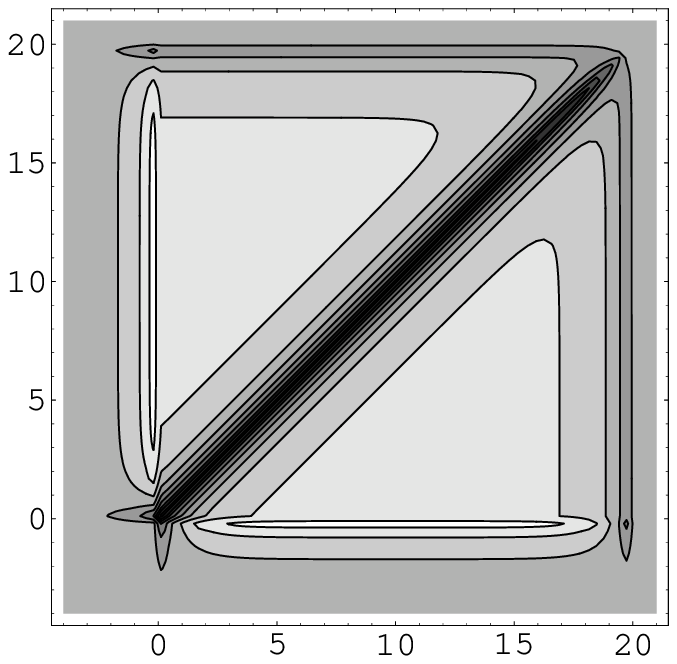}}}
\put(250,0){\makebox(240,240){\includegraphics{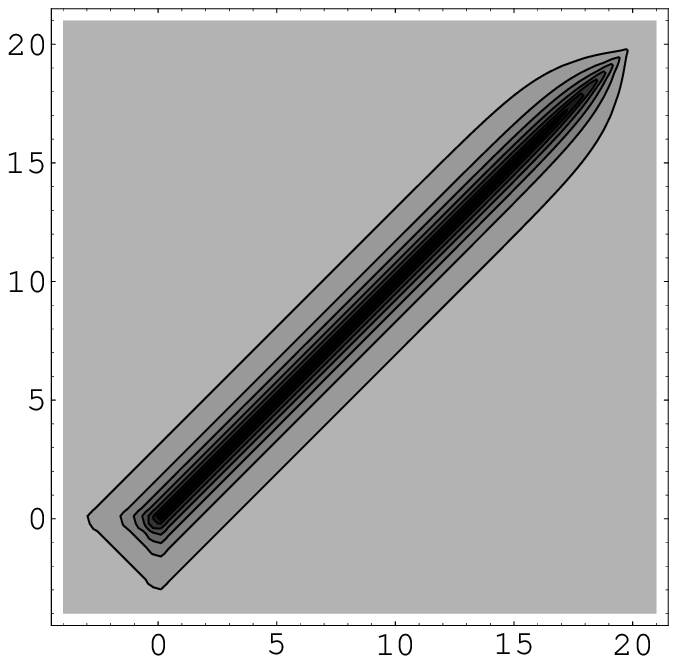}}}
\put(130,0){\makebox(20,20){\large $\frac{\Gamma}{c} x_1$}}
\put(370,0){\makebox(20,20){\large $\frac{\Gamma}{c} x_1$}}
\put(10,110){\makebox(20,20){\large $\frac{\Gamma}{c} x_2$}}
\put(250,110){\makebox(20,20){\large $\frac{\Gamma}{c} x_2$}}
\put(30,217){\makebox(148,20){\large (a) total output $\psi_{\mbox{out}}$}}
\put(280,217){\makebox(170,20){\large(b) nonlinear part 
$\Delta \psi_{\mbox{nonlin.}}$}}

\end{picture}
\caption{\label{output} Contour plots of (a) the output 
wavefunction $\psi_{\mbox{out}}(x_1,x_2)$ and 
(b) the nonlinear component 
$\Delta \psi_{\mbox{nonlin.}}(x_1,x_2)$ of the output 
for a resonant rectangular
input wavepacket of length $L=20 c/\Gamma$. The contour shading
corresponds to amplitudes ranging from $-4/L$ for black to $+2/L$
for white. The dark grey shading at the edges of the graphs 
correspond to zero amplitude. The light grey shading of
the triangular plateu regions in (a) correspond to an amplitude of
$1/L$ equal to the input amplitude of the rectangular wavepacket.
}
\end{figure}

Since the absorption of a photon is strongest at resonance,
a resonant input should also produce the strongest nonlinear
effect in the field-atom interaction. 
In order to investigate this resonant nonlinearity, 
we consider the response to a rectangular input wavepacket
given by
\begin{equation}
\phi_{\mbox{in}}(x)=\left \{ 
\begin{array}{cl}
\frac{1}{\sqrt{L}} & 
\mbox{for}\hspace{0.2cm} 0<x<L
\\[0.2cm]
0 & \mbox{else}.
\end{array}
\right. 
\end{equation}
The linear and nonlinear parts of the output wavefunction 
for this rectangular wavepacket can be determined analytically.
They read
\begin{equation}
\phi_{\mbox{out}}(x) = \left \{ 
\begin{array}{cl}
-\frac{2}{\sqrt{L}} \left(1-\exp(-\frac{\Gamma L}{c})\right) 
\exp(\frac{\Gamma}{c} x) & 
\mbox{for}\hspace{0.2cm} x<0
\\[0.2cm] 
-\frac{1}{\sqrt{L}} \left(1-2 \exp(-\frac{\Gamma}{c}(L-x))\right) & 
\mbox{for}\hspace{0.2cm} 0<x<L
\\[0.2cm]
0 & \mbox{else}
\end{array}
\right.
\end{equation}
and
\begin{equation}
\Delta\psi_{\mbox{nonlin.}}(x_1,x_2) = \left \{ 
\begin{array}{cl}
-\frac{4}{L} \left(1-\exp(-\frac{\Gamma L}{c})\right)^2 
\exp(\frac{\Gamma}{c} (x_1+x_2-2 
\mbox{Max}\{0,x_1,x_2\})) 
& 
\mbox{for}\hspace{0.2cm} x_i<L 
\\[0.2cm] 
0 & \mbox{else}.
\end{array}
\right.
\end{equation}
Figure \ref{output} shows the output wavefunction 
$\psi_{\mbox{out}}$ and
the nonlinear component $\Delta \psi_{\mbox{nonlin.}}$
at an input pulse length of $L=20 c/\Gamma$. The most
remarkable feature of the nonlinear contribution is
its localization around $x_1=x_2$. This is a direct
consequence of the local interaction between the 
two photons. 

A detailed discussion of the two time 
correlation originating from this spatiotemporal
locality of the interaction is given elsewhere 
\cite{Koj02}. In the present paper, we focus on the coherent
properties of the two photon wavefunction.
For this purpose it is useful to simplify the results by assuming
the limit of long pulses, $L\gg c/\Gamma$, and concentrating on
the region within the pulse, $0<x_i<L$. In this limit, the
photon-photon interaction becomes independent of the pulse
shape effects caused by the sudden rise and fall of the 
rectangular pulse amplitude. The results should then
apply to any pulse with an input amplitude varying slowly
on a scale of $c/\Gamma$, where the pulse length parameter
$L$ defines the local photon density as $2/L$.
The output amplitudes are then given by
\begin{equation}
\phi_{\mbox{out}}(x_1)\phi_{\mbox{out}}(x_2) = \frac{1}{L}
\end{equation}
\begin{equation}
\label{eq:psinl}
\Delta\psi_{\mbox{nonlin.}}(x_1,x_2) = -\frac{4}{L}
\exp(-\frac{\Gamma}{c} |x_1-x_2|)
\end{equation}
\begin{equation}
\psi_{\mbox{out}}(x_1,x_2) = \frac{1}{L}
\left(1- 4 \exp(-\frac{\Gamma}{c} |x_1-x_2|)\right).
\end{equation}
In the long pulse limit, the linear part of the output 
wavefunction is nearly equal to the original input pulse. 
However, the nonlinear contribution reduces this overlap
by scattering photons into other modes according to
\begin{eqnarray}
\langle \psi_{\mbox{out}}\mid \psi_{\mbox{in}}\rangle
&=& 1 + \int dx_1 dx_2 \; \psi^*_{\mbox{out}}(x_1,x_2)
\Delta \psi_{\mbox{nonlin}}(x_1,x_2).
\nonumber \\
&\approx& 1 - \frac{4}{L} \int dx_- 
\exp(-\frac{\Gamma}{c} |x_-|) \;=\; 1 - \frac{8 c}{\Gamma L}.
\end{eqnarray}
The probability that the two photons will be scattered out
of the input mode is therefore approximately equal to 
\begin{equation}
1-|\langle \psi_{\mbox{out}}\mid \psi_{\mbox{in}}\rangle|^2
\approx \frac{16 c}{\Gamma L}.
\end{equation}
The long pulse limit requires that this
fraction is never close to one. However, the result can be
used to define a scattering cross section for the two photons.
If we think of the first photon as being in a random position
within the pulse, the chance of finding the second photon within
a distance $\leq \sigma$ should be equal to $2\sigma/L$.
The interaction cross section $\sigma$ for the two photon
nonlinearity can then be defined as $\sigma = 8 c/\Gamma$.
Note that $c/\Gamma$ is the coherence length of spontaneous
emission from the atom. The nonlinear photon-photon interaction
mediated by the two level atom therefore appears to extend over 
a region eight times longer than this coherence length. 

\section{Entanglement and four wave mixing in the nonlinear
component}
\begin{figure}
\begin{picture}(320,200)
\put(60,20){\makebox(250,150){\includegraphics{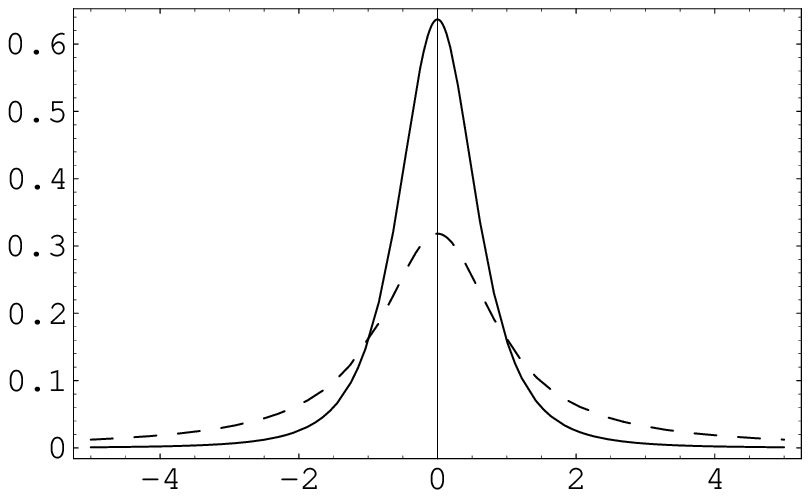}}}
\put(150,170){\makebox(100,20){\large $I_{\mbox{scatter}}(k)$}}
\put(10,100){\makebox(50,20){\large 
$\frac{\Gamma^2 L}{16 c^2}\;I_{\mbox{\small scatter}}$}}
\put(180,0){\makebox(30,20){\large $\frac{c}{\Gamma}\; k$}}
\end{picture}
\caption{\label{spect} Frequency spectrum $I_{\mbox{scatter}}$
of the photons in the nonlinear component 
$\Delta \psi_{\mbox{nonlin.}}$. The intensity $I$ and the 
frequency $k$ have been scaled in such a way that the area
of the spectral line in the graph is equal to one. 
The dashed line shows the Lorentzian line of spontaneous
emission from the two level atom derived from the same
model \cite{Hof95}. Note that the area of this line is
also one. The comparison shows that the spectrum of photons
scatters by spontaneous four wave mixing at the single
atom is narrower than the spectrum of spontaneous emission.
}
\end{figure}

In the long pulse limit, the input mode is very nearly a
plane wave resonant with the two level atom ($k=0$).
It is therefore possible to describe the scattering effect
as a four wave mixing effect changing the photon frequencies
from $k_0=0$ to $+k$ and $-k$, respectively.
The k-space representation of the output wavepacket can be
obtained by using the local Fourier transform in the
spatial region from $x_i=0$ to $x_i=L$ given by
\begin{eqnarray}
\label{eq:krep}
\psi_{\mbox{out}}(k_1,k_2) &=& \frac{1}{L}
\int_0^L dx_1 dx_2 \exp(-i k_1 x_1)\exp(-i k_2 x_2)
\psi_{\mbox{out}}(x_1,x_2)
\nonumber \\ &\approx&
\delta_{k_1,0}\delta_{k_2,0}
-\frac{8 \Gamma c}{L(\Gamma^2+c^2 k_1^2)} \delta_{k_1,-k_2},
\end{eqnarray}
where $k_i$ can have values equal to integer multiples of
$2\pi/L$. Note that this discretization of $k_i$ is 
necessary to preserve the correct normalization of the 
quantum state. 
The phase matching conditions of four wave mixing is
expressed in equation (\ref{eq:krep}) as a Kronecker
delta, $\delta_{k_1,-k_2}$, ensuring that the sum of
$k_1$ and $k_2$ is indeed zero. As a result of this
strong correlation between $k_1$ and $k_2$, the k-space
representation of the two photon output is the
Schmidt decomposition of the entangled state \cite{Nie2},
\begin{equation}
\label{eq:schmidt}
\mid \psi_{\mbox{out}}\rangle = 
\mid k_1=0;k_2=0\rangle - 
\sum_{k}\frac{8 \Gamma c}{L(\Gamma^2+c^2 k^2)}
\mid k;-k\rangle. 
\end{equation}
According to this representation of the two photon state,
the single photon density matrix can be written as a
mixture of k-eigenstates with
\begin{equation}
\hat{\rho} = 
\left(1 - \frac{16 c}{\Gamma L}\right) \mid k=0\rangle
\langle k=0\mid 
+ \sum_k  \left(\frac{8 \Gamma c}{L(\Gamma^2+c^2 k^2)}\right)^2
\mid k \rangle
\langle k \mid. 
\end{equation}
This density matrix defines the single photon coherence of
the output. In particular, the frequency spectrum of the
scattered light is given by a squared Lorentzian,
\begin{eqnarray}
I_{\mbox{scatter}}(k) &=& 
\frac{1}{\Delta k} \langle k \mid \hat{\rho} \mid k \rangle
\nonumber \\
&=&
\frac{16 c}{\Gamma L} \; 
\frac{2 c \Gamma^3}{\pi (\Gamma^2+c^2 k^2)^2}.
\end{eqnarray}
Note that the resolution factor $\Delta k = 2\pi/L$ is 
required to adjust the normalization of the continuous
spectrum $I_{\mbox{scatter}}(k)$ to the discrete 
distribution given by $\hat{\rho}$. 
Figure \ref{spect} shows this scattering spectrum 
in comparison with the spontaneous emission spectrum
of the two level atom. It should be noted that the
squared Lorentzian of the scattering spectrum is
narrower than the Lorentzian of spontaneous emission.
This spectral feature clearly distinguishes the two
photon scattering process from an incoherent sequence
of absorption and reemission and may serve as an 
indication of spontaneous four wave mixing in experiments
where low detection efficiencies prevent an evaluation
of two photon coincidences.

As this analysis shows, the resonant nonlinear interaction 
of the two photons at the atom causes correlated changes in
the frequencies of the photons.
Since the output state is completely quantum coherent, 
the noise in the single photon density matrix
actually indicates entanglement between the scattered photons.
This situation is quite similar to the creation of photon 
pairs by spontaneous parametric downconversion. 
In fact, it may also be possible to create entangled photon 
pairs from the spontaneous four wave mixing effect at a
single atom nonlinearity by isolating the nonlinear part 
of the two photon response to a coherent input field
through destructive interference with an appropriate 
reference pulse. This method will be discussed in the
next section.

\section{Generation of photon pairs using coherent input light}

Spontaneous four wave mixing can only occur if two photons
interact. Moreover, phase matching requires that a photon
scattered to $+k$ must always be accompanied by a photon
scattered to $-k$. It is therefore possible to use the 
nonlinear photon-photon interaction to generate correlated
photon pairs from a coherent input pulse by selecting
the corresponding output ports in a spectrometer. 
However, even better results for photon pair creation
may be achieved 
if the linear component is removed by interference with
another coherent light field using a method similar
to the one applied to parametric downconversion in
\cite{Lu02}. 

For any pulse shape defined by the wavefunction $\phi$,
it is possible to define a weak coherent state $\mid
\alpha \rangle$ with a low average photon number 
$|\alpha|^2\ll 1$. This coherent state can then be expanded
into components with zero, one, and two photons.
Using $\mid \mbox{Vac.}\rangle$ for the vacuum state,
$\mid \phi \rangle $ for the single photon pulse, and
$\mid \phi;\phi \rangle $ for the two photon pulse,
this expansion reads
\begin{equation}
\mid \alpha \rangle \approx \mid \mbox{Vac.}\rangle
+ \alpha \mid \phi \rangle +\frac{\alpha^2}{\sqrt{2}}
\mid \phi; \phi \rangle + \ldots 
\end{equation}
The unitary operator $\hat{U}$ describing the response
of the two level system can now be applied separately
to the vacuum, to the single photon state, and
to the two photon state. 
The vacuum state is not changed by the interaction at all
($\hat{U} \mid \mbox{Vac.}\rangle = 
\mid \mbox{Vac.}\rangle$). In the resonant long pulse limit, 
the single photon component changes its phase by $\pi$, but
remains nearly unchanged otherwise. However, the two photon
component is changed by the addition of 
$\mid \Delta \psi_{\mbox{nonlin.}}\rangle$.
The expansion of the output state therefore reads
\begin{eqnarray}
\mid \psi_{\mbox{out}}\rangle &\approx&
\mid \mbox{Vac.}\rangle
- \alpha \mid \phi \rangle +\frac{\alpha^2}{\sqrt{2}}
\mid \phi; \phi \rangle + \frac{\alpha^2}{\sqrt{2}}
\mid \Delta \psi_{\mbox{nonlin.}}\rangle
\nonumber \\
&\approx& \mid -\alpha \rangle + \frac{\alpha^2}{\sqrt{2}}
\mid \Delta \psi_{\mbox{nonlin.}}\rangle.
\end{eqnarray}
The linear component can therefore be represented by the
weak coherent state $\mid -\alpha \rangle$ with the same
coherence properties as the original pulse. 
This coherent pulse can be removed by destructive 
interference with a much stronger reference pulse of the
same shape at a high reflectivity beam splitter.
Note that the high reflectivity of the beam splitter is
necessary to avoid quantum noise effects in the interaction
that would appear as photon losses in the final output. 
If these conditions are met, the destructive interference
may be represented by the displacement operator 
$\hat{D}(\alpha) = \exp(\alpha \; \hat{a}^\dagger 
- \alpha^* \hat{a})$. 
For $|\alpha| \ll 1$, this operator is only slightly 
different from $\hat{1}$, but it does have the fundamental
property that $\hat{D}(\alpha)\mid -\alpha \rangle
=\mid \mbox{Vac.} \rangle$. The final output therefore
reads
\begin{eqnarray}
\hat{D}(\alpha) \mid \psi_{\mbox{out}}\rangle &\approx&
\hat{D}(\alpha) \mid -\alpha \rangle +
\hat{D}(\alpha) \frac{\alpha^2}{\sqrt{2}}
\mid \Delta \psi_{\mbox{nonlin.}}\rangle
\nonumber \\ &\approx&
\mid \mbox{Vac.} \rangle + \frac{\alpha^2}{\sqrt{2}} 
\mid \Delta \psi_{\mbox{nonlin.}}\rangle.
\end{eqnarray}
This output wavefunction now contains only a zero and
a two photon component. The one photon component has been
eliminated by the interference effects at the high 
reflectivity beam splitter.
It is therefore possible to generate entangled photon pairs
with a two photon wavefunction described by 
$\Delta \psi_{\mbox{nonlin.}}$ using a coherently driven 
dissipation free two level atom and an interferometric setup.
The average number of photon pairs created in each pulse is
then given by
\begin{equation}
\frac{|\alpha|^4}{2} 
\langle \Delta \psi_{\mbox{nonlin.}}
\mid \Delta \psi_{\mbox{nonlin.}}\rangle
= \frac{8c}{\Gamma L} |\alpha|^4  
.
\end{equation}
In the long pulse limit, it is possible to approximate
continuous input light as a sequence of rectangular pulses
of length $L\gg c/\Gamma$. The intensity of the pump 
light is then given by $I_{\mbox{in}}=c|\alpha|^2/L$ 
and the rate of pair creation $R_{\mbox{pair}}$ is 
given by the average number of pairs per pulse divided
by the pulse duration $L/c$. The result of this estimate
reads
\begin{equation}
R_{\mbox{pair}} = \frac{8}{\Gamma} I_{\mbox{in}}^2,
\end{equation}
where higher order many photon effects are negligible if
$I_{\mbox{in}}\ll \Gamma$. The pair creation rate is
therefore also limited to $R_{\mbox{pair}} \ll \Gamma$.
However, $\Gamma$ is usually in the range of nanoseconds,
so considerable pair rates should be possible.

According to equation (\ref{eq:schmidt}), the quantum state 
of the emitted photon pair can be written as 
\begin{equation}
\mid \Delta \psi_{\mbox{nonlin.}}\rangle 
= - \sum_{k}\frac{8 \Gamma c}{L(\Gamma^2+c^2 k^2)}
\mid k;-k\rangle,
\end{equation}
where $k$ represents the discretized k-space with
$\Delta k=2\pi/L$. 
In real space representation, the same entanglement is
expressed by the coefficients $\Delta \psi_{\mbox{nonlin.}}
(x_1,x_2)$ given by equation (\ref{eq:psinl}). 
These representations show the same time-frequency correlations
as a phase matched parametric downconversion, that is
$x_1\approx x_2$ and $k_1 = - k_2$.
It may therefore be possible to use photon pairs created by
four wave mixing in applications similar to those of
downconverted photons. 

\section{Conclusions}
We have described the spatiotemporal dynamics of a one
dimensional light field interacting with a single two level
atom for input states with up to two photons.
In the case of a resonant two photon input, 
the interaction at the atom results in spontaneous four 
wave mixing effects,
scattering the photons to higher and lower frequencies.
Since this scattering effect is fully quantum coherent,
the resulting output state is entangled in frequency 
and time. 

For a coherent state input, it is possible to remove the
linear single photon and two photon components by destructive
interference with a reference pulse.
The remaining output then consists of the vacuum state and
a small contribution from the nonlinear two photon component.
This output is very similar to the output from spontaneous
parametric downconversion. It may therefore be possible to 
realize a source of entangled photon pairs using the 
spontaneous four wave mixing effects at a single two level
atom.



\end{document}